\documentstyle[aps,preprint]{revtex}
\smallskip
\begin{document}
\newcommand{\bc}{\begin{center}}
\newcommand{\ec}{\end{center}}
\newcommand{\be}{\begin{equation}}
\newcommand{\ee}{\end{equation}}
\def\e{{\rm e}}
\title{Temperature dependence of local states due to $S=1/2$ impurities and
their correlation in a $S=1$ Heisenberg chain}
\author{Pascal Roos and Seiji Miyashita}
\address{Department of Earth and Space Science, Osaka University, Toyonaka,
Osaka 560-0043}
\date{\today}
\maketitle
\begin{abstract}
We study the temperature dependence of the low temperature spin configurations,
investigating the magnetization profile of the local states due to the
impurities and
the two point correlation function centered in one of the impurities.
This correlation is found to be weak against temperature effects although the
magnetization profile in the triplet state is visible up to higher
temperatures.
Here we introduce  a loop cluster quantum Monte-Carlo method with a fixed
magnetization $M_z$ in order to study the
correlations in the ground state of a given value of $M_z$.
From the population distribution of magnetization, the very small energy gap
between the quasi degenerate states due to the impurities is obtained.
\end{abstract}

\section{Introduction}
Since the Haldane conjecture\cite{Haldane}, the peculiar properties of 
quantum spin systems have been interested in.
In particular, properties of integer spin antiferromagnetic quantum 
spin chain have been studied in detail: 
theoretically \cite{Affleck-Houches,Affleck}, 
numerically\cite{Botet,Kennedy,White,Miyashita,Yamamoto} and 
experimentally\cite{Buyers,Renard,Katsumata}. 
There, important concepts of Valence Bond Solid states (VBS) \cite{Affleck}  
and hidden order parameter \cite{Den Nijs,Oshikawa} have been introduced.
Studies involving the impurity (non-magnetic impurities and/or magnetic
impurities) on spin chains have been also a current 
topic\cite{Hagiwara,Kaburagi,Sorensen,WWang,XWang}.

As far as we know, only few of them deal with these impurity problem at
finite temperatures.
Hence, in the present article we will focus on temperature dependence of
local states due
to $S=1/2$ impurities and their correlation in a $S=1$ Heisenberg chain.
Here we
mainly use the loop cluster quantum Monte-Carlo method (LCQMC).
We will be concerned here with the problem of two $S=1/2$ spins
embedded in a $S=1$
antiferromagnetic Heisenberg periodic chains.
While an impurity causes a doublet state\cite{Hagiwara}, two impurities
cause a structure with
singlet and triplet states similar to the case of the Kennedy triplet in
the edge effects\cite{Kennedy,Miyashita}.
This impurity-induced ground state presents a localized structure
around the impurities
and it has an inhomogeneous order.

In the next section, we give a quick survey of the LCQMC method used to
simulate our system.
In Sect. \ref{sec:eg}, we obtain the energy gap between quasi degenerate ground
state making use of the population distribution of magnetization.
In Sect. \ref{sec:mp}, we investigate magnetization profile of the impurity system.
In Sect. \ref{sec:cf}, temperature dependence of the correlation function 
is investigated and is compared with the magnetization profile.
Sect. \ref{sec:dd} is devoted to the summary and discussion where we
also discuss the metastability of local states making use of snapshot
obtained in a world-line updating algorithm..

\section{Model and Method}
\label{sec:mm}
The model we deal with in this paper is the one dimensional
antiferromagnetic Heisenberg  $S=1$ system where we replace some of
the spins by $S=1/2$ impurities. 
The general hamiltonian of this model
is written as
\be
{\cal H}=J\sum_{i}\vec{S}_i\vec{S}_{i+1}
+J'\sum_{i'}(\vec{S}_{i'-1}\vec{\sigma}_{i'}+
\vec{\sigma}_{i'}\vec{S}_{i'+1}),	
\label{ham}
\ee
where the first summation concerns the bonds of $S=1$ spins and the
second one the contribution
due to the impurities. Here $S$ (resp. $\sigma$) represents spin one (resp.
one-half).
Since we are interested in a purely antiferromagnetic case, $J$ and $J'$
are both positive and we adopt the periodic boundary condition.

In this paper we consider the lattice with two impurities specifically.
The impurities are located on sites ensuring a maximum severing compatible
with the periodic boundary condition, and the ratio $J'/J$ is fixed to unity. 
For this particular impurity location,
the ground state is expected to be a triplet $S=1$, with the first exited
state of the singlet $S=0$ just above it, 
with a small gap $\Delta_1$ between them. 
This behavior is similar to the edges effects which happen in a pure spin
one chain with open boundary conditions.
This energy gap corresponds to the gap between the Kennedy triplet 
and the singlet state in the open chain and is exponentially small
with $L$\cite{Kennedy,Miyashita}. Thus even at $T=0.01J$, $\Delta_1\ll T$
for long chains such as $L=64$.
There is also another gap between the ground state and the excitation continuum,
which we will refer as the Haldane gap $\Delta_H$ henceforth: 
$\Delta_H\simeq0.4J$ in the pure $S=1$ model.
The expected low energy spectrum is presented in Fig.~\ref{fig:energy}.
If we shift one of the impurities by one site, the singlet
state becomes the ground state and the triplet states locate above it,
although the global excitation spectrum is similar to Fig.~\ref{fig:energy}.

In order to obtain the finite temperature properties we mainly use
the Loop Cluster Quantum Monte-Carlo (LCQMC) method \cite{lcqmc}.
The LCQMC allows to perform the continuous time
calculation easily where we can avoid the Trotter number extrapolation
in the traditional World-line Quantum Monte-Carlo (WLQMC) method.
The configuration update is done in the LCQMC without conserving
the total magnetization or the winding number, on the contrary to the WLQMC.
In general, LCQMC releases us from strong autocorrelation and thus allow us
to reach to the equilibrium state at very low temperatures. 
We compared the simulation of LCQMC and a traditional WLQMC.
We found that about $10^5$ MCS gives good results in the present study by
LCQMC, while more than $10^6$ are necessary to obtain the results with the
same accuracy by WLQMC at $T=0.1J$.

However, in a simple minded LCQMC we can not specify the value of
the magnetization $M_z$ and if some states are nearly degenerate around the
ground state, LCQMC always equilibrates among these states.
Thus, it is difficult to obtain the true ground state configurations.
In WLQMC we can control the value of $M_z$ in the initial state and
keep it by suppressing the global flip which changes the magnetization.

This nearly degenerate ground state happens in the present study.
Thus it is preferable to control $M_z$ in LCQMC.
In this circumstance we introduced a method where we can specify the value
of $M_z$ ($M_z$-specifying LCQMC) in order to avoid this difficulty.
There are two ways to control $M_z$ in LCQMC.
One is the way where we stop the flip of the clusters of nonzero
magnetization. Thus the magnetization in the initial state is kept.
The other is the way where we perform standard LCQMC and store 
the data separately according to $M_z$.
If we need the information for a specific value of $M_z$, we
use the data with that value of $M_z$ only.

Here, we adopted the latter method. In order to check the method we compare
the low temperature results obtained by the LCQMC and those obtained by 
Exact Diagonalization (ED).
Since ED is restricted to rather short chains,
we compared the result for a chain length of $L=12$. 

In $L=12$ the gap $\Delta_1$ is rather big due to finite size effects and
is about $0.6174J$. Thus, we can study the
ground state at $T\ll\Delta_1$. For example $T=0.01J$ is quite low enough.
First we compare data within $M_z=1$ obtained by ED and the data obtained
by the above mentioned method.

We show the profile of magnetization:
\begin{equation}
\{M^z(i)\}= \{\langle \sigma_i\rangle {\rm or}\ \langle S_i\rangle\}
\end{equation}
for $L=12$ with $M_z=1$ obtained by the both methods in Fig. \ref{fig:cmag}.
Here we find that the method works very well.

Next, we compare data without specifying magnetization, where we
need to sum up the data obtained by ED of the ground states for $M_z=-1$,
0 and 1 in order to compare the data .
If we do not specify the magnetization, trivially we have $M^z(i)=0$.
Thus here we compare the correlation function: 
\begin{equation}
\{C_j=\langle \sigma_i S_j\rangle\},
\end{equation}
where we observe the correlation between a $S=1/2$ spin $\sigma_i$ 
and other spins. 
In Fig. \ref{fig:ccor}(a) we show the data without specification of $M_z$ obtained by
ED and LCQMC, respectively.
Here we again find that the method works very well.
We also compare data of the correlation function within $M_z=1$ in
Fig. \ref{fig:ccor}(b).
Thus we conclude the method to specify the value of $M_z$ is valid and works
practically.

\section{Energy gap of the nearly degenerate ground states}
\label{sec:eg}
We can utilize the fact that the equilibrium among nearly degenerate states
is rather easily realized in LCQMC, which is difficult in WLQMC.
In the equilibrium state, the probability of the triplet state and the
probability of the singlet state in the quasi degenerate states are given,
\begin{equation}
p_{\rm t}=\e^{-{E_G\over k_{\rm B}T}}/Z\quad {\rm and}\quad
p_{\rm s}=\e^{-{E_G+\Delta_1\over k_{\rm B}T}}/Z,
\end{equation}
respectively. Here $Z$ is the partition function.
Because the energies of other states are much higher, i.e.
the Haldane gap $\Delta_{\rm H}$ is much larger than the temperature:
$\Delta_{\rm H}\gg T\gg\Delta_1$,
we have 
\begin{equation}
p_{\rm t}={1\over 3+\e^{-{\Delta_1\over k_{\rm B}T}}}\quad {\rm and} \quad
p_{\rm s}={\e^{-{\Delta_1\over k_{\rm B}T}}
\over 3+\e^{-{\Delta_1\over k_{\rm B}T}}}.
\end{equation}
Thus by counting the number of the Monte Carlo steps (MCS) 
for $M_z=-1,0$ and 1, say $N_{-1},N_0$ and $N_1$, respectively,
we can estimate the $\Delta_1$ from the relation: 
\begin{equation}
r= {N_{-1}+N_1-N_0\over N_{-1}+N_1+N_0}=
{1-\e^{-{\Delta_1\over k_{\rm B}T}} \over 3+\e^{-{\Delta_1\over k_{\rm B}T}}}.
\end{equation}
In Fig. \ref{fig:dist}
we show the histograms of $\{ N_M(T)\}$ for $T=0.01J$ and $T=0.2J$.
At $T=0.01J$ we find only $M_z=\pm 1$ and $0$ which come from the 
nearly degenerate four states. On the other hand at $T=0.2J$ the
magnetization distributes up to higher values. In 
table~\ref{tab:NvsT} we show the value of $\{ N_{M_z}\}$ for different
temperatures.
In order to obtain the gap, in Fig. \ref{fig:gap} we plot the gap $\Delta_1$  
obtained by
\begin{equation}
-\ln\left({1-3r\over3+r}\right)={\Delta_1\over k_{\rm B}T}
\label{gap1}
\end{equation}
as function of $k_{\rm B}T$.
Here we estimate $\Delta_1\simeq 0.0047J\pm 0.0002$. Here the error 
bar is obtained as the standard deviation in the values of $\Delta_1$ 
estimated at temperatures $T \leq 0.03J$.
Although it has been so far difficult to obtain this gap because of the smallness,
now we have a method to estimate it, which is one of the advantages of the 
LCQMC.

\section{Magnetization profile}
\label{sec:mp}
Because a doping with S=1/2 causes the local states with rather widely spreaded
impurity-induced magnetization, it is expected that the local states of
doped spins correlate
each other and cause a quasi-long range order in the system.
In order to see this correlation of local states, we investigate the
magnetization profile in
the ground state. By LCQMC, we obtain the profile at $T=0.01J$.
In Fig. \ref{fig:mag} we show the magnetization profile for $L=64$
with $M_z=1$.
Here, the impurities are located on the site 16 and 48.
This figure implies the presence of impurity induced Long Range Order
(LRO). It is a
ferromagnetic LRO of the effective spins of the local states while it is an
inhomogeneous antiferromagnetic order of the original spins.

If we study the magnetization profile in the subspace of $M_z=1$,
the local structure should remain till a rather high temperature because
of the gap in this subspace is about $\Delta_H$.
At high temperatures the correlation between them collapses.
In order to see the degree of this correlation, we observe the staggered 
magnetization:
\be
M_{\rm SG}=\sum_{i=1}^L (-1)^iS_i^z.
\label{stgmg}
\ee
In Fig. \ref{fig:stgmag}, the temperature dependence of
$\langle M_{\rm SG}\rangle$ is shown.
Here, we find that the correlation persists until $T_{1}\simeq 0.4J$.
Even above this temperature we find the local structure remains, but the
magnetization of the local state fluctuates in time and thus
$\langle M_{\rm SG} \rangle $ vanishes.
The local structure persists until $T_{2}\simeq 1J$. 

In order to investigate the true equilibrium state we have to study the
system without specifying the value of $M_z$. However,
if we do not fix the magnetization we do not find any significant profile,
that is $<S_i>=0$ in principle because of the degeneracy of
$M_z=\pm 1$ and $M_z=0$, 
even if there exists some steady magnetization profile in each of them.
Thus we should investigate the spin correlation function instead of 
the magnetization profile in order to
study whether an intrinsic correlation exists or not.

\section{Correlation function}
\label{sec:cf}
As pointed out in the previous section, the LRO in the ground state can
be viewed as the correlation between the local states. 
In order to explore the correlation behavior, notably the
temperature dependence, we investigate the spatial profile of the two point
correlation function $<S_iS_j>$.

Some results for $L=12$ has been shown in Sect. \ref{sec:mm}.
In this section we study a system of $L=64$.
In Fig. \ref{fig:corlT}(a), the correlation function in the $M_z=1$ subspace
at $T=0.01J$ is shown which represents the ground state configuration.
It should be noted that the correlation appears in rather small amount in
comparison with the magnetization profile, which can be explained by the
trivial relation $|<S_iS_j>| \:\ge\: |<S_i><S_j>|$.

The correlation function without specifying $M_z$ value 
is presented in Fig. \ref{fig:corlT}(b). 
Although the local structure around the impurity remains, 
the correlation function between the two local structures is much reduced.
This reduction can be understood from the argument in Sect. \ref{sec:eg}.
In the true ground state there are two parallel and
one anti-parallel configurations of the local structure. Thus we expects that
the correlation between the local states, which corresponds to the values
around $i=48$, is one third of that in Fig. \ref{fig:corlT}(a).
However the temperature is higher than the gap $\Delta_1$, and thus the 
singlet (excited) state also contributes to the correlation. 
Thus as we discussed in Sect. \ref{sec:eg}, 
value of the correlation function, $C$, is estimated as
\be
C=(\frac{1-e^{{-\Delta_1\over k_{\rm B}T}}}{3+e^{{-\Delta_1\over k_{\rm B}T}}})C_0 ,
\ee
where $C_0$ is the corresponding value in Fig. \ref{fig:corlT}(a).
The reduction ratio
$(1-e^{{-\Delta_1/k_{\rm B}T}})/(3+e^{{-\Delta_1/k_{\rm B}T}})$
has been  estimated in the Sect. III to be $0.105$. In Fig. 8(b) at $T=0.01J$
the ratio is found to be consistent and we still find some correlation.
If the temperature increases up to $T=0.05J$, the reduction rate is $0.017$
and the correlation can not be seen any more. (Fig. \ref{fig:corlT}(c))

On the other hand if we fix the magnetization in $M_z=1$ the 
correlation is only gradually reduced as the temperature increases.
In Fig. \ref{fig:corT}(a) we show the temperature dependence of this case.
In Fig. \ref{fig:corT}(b) the summation of the staggered correlation
\be
C_{\rm SG}=\sum_{i=1}^L (-1)^{|i-j|}S_i^zS_j^z.
\ee
is shown. Here $C_{\rm SG}$ reduces to half around $T=0.5J$ which corresponds
to $T_1$ where the correlation between the local state vanishes.
The correlation within a local state remains, which results nonzero
value of $C_{\rm SG}$ above $T_1$. 
$C_{\rm SG}$ begins to grow significantly around $T=1J$ which corresponds to 
$T_2$ in Sect. \ref{sec:mm} where the local state is formed.

\section{Summary and Discussion}
\label{sec:dd}

We have investigated the correlation between the impurity induced 
local magnetic states in $S=1$ antiferromagnetic chain. 
By investigating the equilibrium distribution of $M_z$ at very low temperatures,
which becomes possible in LCQMC, we find a new method to estimate
the energy gap between nearly degenerate states.
We have studied how the impurity-induced correlation function, that is to say,
the correlation between the local states around the impurities, decays when the
temperature increases.
We found that the correlation is very weak against the temperature
although it is robust in the ground state.
The correlation survives only below the temperature of the order of $\Delta_1$
which is the gap between the quasi-degenerate ground state due to the impurity.

Finally we would like to point out the metastable nature of the 
local state. Because the local state is rather tightly bounded,
it tends to move collectively. Thus we expect that
the dynamics of the total motion of the cluster is rather slow.
If we study the spin configuration in WLQMC we can find rather
stable magnetization profile even at high temperatures and also
in $M_z=0$ subspace, which should be a short time metastable state. 
In Fig. \ref{fig:discmag} we show an example of such configuration.
In LCQMC the update is very rapid and we 
can not see such metastable state. If we are interested in the 
relaxation phenomena after changing parameters of the system, such 
as switching off the magnetization, etc., the dynamics of such
metastable states becomes important.
While at high temperatures they relax by the temperature effect,
they would relax through quantum tunneling at low temperatures.
The quantum mechanical life time of the metastability can be 
estimated by the correlation 
length through the Trotter axis, as has been investigated as the local 
susceptibility in the studies of quantum 
Griffiths-McCoy singularity\cite{Rieger,Ikegami}.
Such dynamics will be studied in the future.

{\bf Acknowledgements:}
The numerical calculations were partly carried in the computer centers 
of Yukawa Institute and Institute of Solid State Physics.
The authors acknowledge Prof.~N.~Kawashima for his kind instructions to
make code for LCQMC and also Prof.~M.~Kaburagi for the use of the package
for exact diagonalization KOBEPACK.
One of the author (P. R.) is supported by the JSPS Research Fellowship for
Young Scientists.

\begin{figure}
\caption{Schematic low energy spectrum}
\label{fig:energy}
\end{figure}

\begin{figure}
\caption{Magnetization profile in $M_z=1$ for ED (circle) and LCQMC at 
$T=0.01J$(square)}
\label{fig:cmag}
\end{figure}

\begin{figure}
\caption{Two point correlation function profile for ED (circle) and LCQMC
at $T=0.01J$ (square) (a) without fixing $M_z$ (b) in $M_z=1$ subspace}
\label{fig:ccor}
\end{figure}

\begin{figure}
\caption{Distribution of $M_z$ for $10^5$ MCS (a)$T=0.01J$ (b)$T=0.2J$}
\label{fig:dist}
\end{figure}

\begin{figure}
\caption{Gap $\Delta_1$ estimated at various temperatures by Eq.(7)}
\label{fig:gap}
\end{figure}

\begin{figure}
\caption{Magnetization profile in $M_z=1$ at $T=0.01J$}
\label{fig:mag}
\end{figure}

\begin{figure}
\caption{Staggered magnetization, Eq.(8), versus $T$ in $M_z=1$}
\label{fig:stgmag}
\end{figure}

\begin{figure}
\caption{Two point correlation function profile
a) in $M_z=1$ at $T=0.01J$ b) without fixing $M_z$ at $T=0.01J$
c) without fixing $M_z$ at $T=0.05J$}
\label{fig:corlT}
\end{figure}

\begin{figure}
\caption{
a) Two point correlation function for various temperatures within $M_z=1$ and
b) summation of the staggered two point correlation function}
\label{fig:corT}
\end{figure}

\begin{figure}
\caption{Magnetization profile in a short time for $M_z=0$ at $T=0.05J$ 
obtained by WLQMC which is expected to be a metastable configuration}
\label{fig:discmag}
\end{figure}

\begin{table}
\caption{Distribution of $\{ N_{M_z}\}$ for various $T$
after $10^5$ Monte-Carlo steps}
\label{tab:NvsT}
\end{table}

\end{document}